\documentclass[twocolumn,aps,superscriptaddress,showpacs]{revtex4}
\usepackage{amsmath,bm}
\usepackage{graphicx}
\usepackage[normalem]{ulem}
\usepackage[dvips]{color}

\begin{document}

\title{Temperature effects on the nuclear symmetry energy and symmetry free
energy with an isospin and momentum dependent interaction}
\author{Jun Xu}
\affiliation{Institute of Theoretical Physics, Shanghai Jiao Tong University, Shanghai
200240, China}
\author{Lie-Wen Chen}
\affiliation{Institute of Theoretical Physics, Shanghai Jiao Tong University, Shanghai
200240, China}
\affiliation{Center of Theoretical Nuclear Physics, National Laboratory of Heavy-Ion
Accelerator, Lanzhou, 730000, China}
\author{Bao-An Li}
\affiliation{Department of Physics, Texas A\&M University-Commerce, Commerce, TX 75429,
and Department of Chemistry and Physics, P.O. Box 419, Arkansas State
University, State University, AR 72467-0419, USA}
\author{Hong-Ru Ma}
\affiliation{Institute of Theoretical Physics, Shanghai Jiao Tong University, Shanghai
200240, China}

\begin{abstract}
Within a self-consistent thermal model using an isospin and momentum
dependent interaction (MDI) constrained by the isospin diffusion data in
heavy-ion collisions, we investigate the temperature dependence of the
symmetry energy $E_{sym}(\rho ,T)$ and symmetry free energy $F_{sym}(\rho
,T) $ for hot, isospin asymmetric nuclear matter. It is shown that the
symmetry energy $E_{sym}(\rho ,T)$ generally decreases with increasing
temperature while the symmetry free energy $F_{sym}(\rho ,T)$ exhibits
opposite temperature dependence. The decrement of the symmetry energy with
temperature is essentially due to the decrement of the potential energy part
of the symmetry energy with temperature. The difference between the symmetry
energy and symmetry free energy is found to be quite small around the
saturation density of nuclear matter. While at very low densities, they
differ significantly from each other. In comparison with the experimental
data of temperature dependent symmetry energy extracted from the isotopic
scaling analysis of intermediate mass fragments (IMF's) in heavy-ion
collisions, the resulting density and temperature dependent symmetry energy $%
E_{sym}(\rho ,T) $ is then used to estimate the average freeze-out density
of the IMF's.
\end{abstract}

\maketitle

\section{Introduction}

The equation of state (EOS) of isospin asymmetric nuclear matter, especially
the nuclear symmetry energy, is essential in understanding not only many
aspects of nuclear physics, but also a number of important issues in
astrophysics \cite%
{ireview98,ibook,bom,diep03,pawel02,lat01,baran05,steiner05,listeiner}.
Information about the symmetry energy at zero temperature is important for
determining ground state properties of exotic nuclei and properties of cold
neutron stars at $\beta $-equilibrium, while the symmetry energy or symmetry
free energy of hot neutron-rich matter is important for understanding the
liquid-gas phase transition of asymmetric nuclear matter, the dynamical
evolution of massive stars and the supernova explosion mechanisms. Heavy-ion
reactions induced by neutron-rich nuclei provide a unique means to
investigate the symmetry energy \cite{ireview98,ibook,baran05}. In
particular, recent analyses of the isospin diffusion data in heavy-ion
reactions \cite{betty04,chen05,lichen05} have already put a stringent
constraint on the symmetry energy of cold neutron-rich matter at sub-normal
densities. On the other hand, the temperature dependence of the symmetry
energy or symmetry free energy for hot neutron-rich matter has received so
far little theoretical attention \cite{chen01,zuo03,lichen06EsymT}.

For finite nuclei at temperatures below about $3$ MeV, the shell structure
and pairing as well as vibrations of nuclear surfaces are important and the
symmetry energy was predicted to increase slightly with the increasing
temperature \cite{don94,dean}. Interestingly, an increase by only about $8\%$
in the symmetry energy in the range of $T$ from $0$ to $1$ MeV was found to
affect appreciably the physics of stellar collapse, especially the
neutralization processes \cite{don94}. At higher temperatures, one expects
the symmetry energy to decrease as the Pauli blocking becomes less important
when the nucleon Fermi surfaces become more diffused at increasingly higher
temperatures \cite{chen01,zuo03,lichen06EsymT}. Based on a simplified
degenerate Fermi gas model at finite temperatures, two of present authors
\cite{lichen06EsymT} have recently studied the temperature dependence of the
symmetry energy and it was shown that the experimentally observed decrease
of the nuclear symmetry energy with the increasing centrality or the
excitation energy in isotopic scaling analyses of heavy-ion reactions can be
well understood analytically within the degenerate Fermi gas model. In
particular, it was argued that the symmetry energy extracted from isotopic
scaling analyses of heavy-ion reactions reflects the symmetry energy of
\emph{bulk nuclear matter} for the emission source. Furthermore, it was
found that the evolution of the symmetry energy extracted from the isotopic
scaling analysis is mainly due to the variation in the freeze-out density
rather than temperature when the fragments are emitted in the reactions
carried out under different conditions.

In the present work, within a self-consistent thermal model using an isospin
and momentum dependent interaction (MDI) constrained by the isospin
diffusion data in heavy-ion collisions, we study systematically the
temperature dependence of the nuclear matter symmetry energy $E_{sym}(\rho
,T)$ and symmetry free energy $F_{sym}(\rho ,T)$. It is shown that the
nuclear matter symmetry energy $E_{sym}(\rho ,T)$ generally decreases with
increasing temperature while the symmetry free energy $F_{sym}(\rho ,T)$
exhibits opposite temperature dependence. The decrement of the symmetry
energy with temperature is essentially due to the decrement of the potential
energy part of the symmetry energy with temperature. Furthermore, the
difference between the nuclear matter symmetry energy $E_{sym}(\rho ,T)$ and
symmetry free energy $F_{sym}(\rho ,T)$ is found to be quite small around
nuclear saturation density. Using the resulting density and temperature
dependent symmetry energy $E_{sym}(\rho ,T)$, we estimate the average
freeze-out density of the fragment emission source based on the measured
temperature dependent symmetry energy from the isotopic scaling analysis in
heavy-ion collisions.

The paper is organized as follows. In Section \ref{EOS}, we introduce the
isospin and momentum dependent MDI interaction and the detailed numerical
method to obtain the EOS of the symmetric nuclear matter and pure neutron
matter at finite temperatures. Results on the temperature dependence of the
symmetry energy and symmetry free energy are presented in Section \ref%
{symmetry energy}. In Section \ref{isoscaling}, we discuss the experimental
data of the isotopic scaling in heavy-ion collisions by means of the
obtained density and temperature dependent symmetry energy. A summary is
given in Section \ref{summary}.

\section{Hot nuclear matter EOS in momentum dependent interaction}

\label{EOS}

Our study is based on a self-consistent thermal model using a modified
finite-range Gogny effective interaction, i.e., the isospin- and
momentum-dependent MDI interaction~\cite{das03}. In the MDI interaction, the
potential energy density $V(\rho ,T,\delta )$ of a thermal equilibrium
asymmetric nuclear matter at total density $\rho $, temperature $T$ and
isospin asymmetry $\delta $ is expressed as follows~\cite{das03,chen05},
\begin{eqnarray}
V(\rho ,T,\delta ) &=&\frac{A_{u}\rho _{n}\rho _{p}}{\rho _{0}}+\frac{A_{l}}{%
2\rho _{0}}(\rho _{n}^{2}+\rho _{p}^{2})  \notag \\
&+&\frac{B}{\sigma +1}\frac{\rho ^{\sigma +1}}{\rho _{0}^{\sigma }}%
(1-x\delta ^{2})  \notag \\
&+&\frac{1}{\rho _{0}}\sum_{\tau ,\tau ^{\prime }}C_{\tau ,\tau ^{\prime
}}\int \int d^{3}pd^{3}p^{\prime }\frac{f_{\tau }(\vec{r},\vec{p})f_{\tau
^{\prime }}(\vec{r},\vec{p}^{\prime })}{1+(\vec{p}-\vec{p}^{\prime
})^{2}/\Lambda ^{2}}.  \label{MDIV}
\end{eqnarray}%
In mean field approximation, Eq. (\ref{MDIV}) leads to the following single
particle potential for a nucleon with momentum $\vec{p}$ and isospin $\tau $
in the thermal equilibrium asymmetric nuclear matter, i.e., \cite%
{das03,chen05}

\begin{eqnarray}
U(\rho ,T,\delta ,\vec{p},\tau ) &=&A_{u}(x)\frac{\rho _{-\tau }}{\rho _{0}}%
+A_{l}(x)\frac{\rho _{\tau }}{\rho _{0}}  \notag \\
&+&B(\frac{\rho }{\rho _{0}})^{\sigma }(1-x\delta ^{2})-8\tau x\frac{B}{%
\sigma +1}\frac{\rho ^{\sigma -1}}{\rho _{0}^{\sigma }}\delta \rho _{-\tau }
\notag \\
&+&\frac{2C_{\tau ,\tau }}{\rho _{0}}\int d^{3}p^{\prime }\frac{f_{\tau }(%
\vec{r},\vec{p}^{\prime })}{1+(\vec{p}-\vec{p}^{\prime })^{2}/\Lambda ^{2}}
\notag \\
&+&\frac{2C_{\tau ,-\tau }}{\rho _{0}}\int d^{3}p^{\prime }\frac{f_{-\tau }(%
\vec{r},\vec{p}^{\prime })}{1+(\vec{p}-\vec{p}^{\prime })^{2}/\Lambda ^{2}}.
\label{MDIU}
\end{eqnarray}%
In the above $\tau =1/2$ ($-1/2$) for neutrons (protons); $\sigma =4/3$; $%
f_{\tau }(\vec{r},\vec{p})$ is the phase space distribution function at
coordinate $\vec{r}$ and momentum $\vec{p}$. The parameters $%
A_{u}(x),A_{l}(x),B,C_{\tau ,\tau },C_{\tau ,-\tau }$ and $\Lambda $ have
been assumed to be temperature independent and are obtained by fitting the
momentum-dependence of $U(\rho ,T=0,\delta ,\vec{p},\tau )$ to that
predicted by the Gogny Hartree-Fock and/or the Brueckner-Hartree-Fock
calculations, the zero temperature saturation properties of symmetric
nuclear matter and the symmetry energy of $31.6$ MeV at normal nuclear
matter density $\rho _{0}=0.16$ fm$^{-3}$ \cite{das03}. The
incompressibility $K_{0}$ of cold symmetric nuclear matter at saturation
density $\rho _{0}$ is set to be $211$ MeV. The parameters $A_{u}(x)$ and $%
A_{l}(x)$ depend on the $x$ parameter according to
\begin{equation}
A_{u}(x)=-95.98-x\frac{2B}{\sigma +1},~A_{l}(x)=-120.57+x\frac{2B}{\sigma +1}%
.
\end{equation}%
The different $x$ values in the MDI interaction are introduced to vary the
density dependence of the nuclear symmetry energy while keeping other
properties of the nuclear equation of state fixed \cite{chen05} and they can
be adjusted to mimic predictions on the density dependence of nuclear matter
symmetry energy by microscopic and/or phenomenological many-body theories.
The last two terms of Eq. (\ref{MDIU}) contain the momentum-dependence of
the single-particle potential. The momentum dependence of the symmetry
potential stems from the different interaction strength parameters $C_{\tau
,-\tau }$ and $C_{\tau ,\tau }$ for a nucleon of isospin $\tau $
interacting, respectively, with unlike and like nucleons in the background
fields. More specifically, we use $C_{\tau ,-\tau }=-103.4$ MeV and $C_{\tau
,\tau }=-11.7$ MeV. We note that the MDI interaction has been extensively
used in the transport model for studying isospin effects in intermediate
energy heavy-ion collisions induced by neutron-rich nuclei \cite%
{li04b,chen04,chen05,lichen05,li05pion,li06,yong061,yong062}. In particular,
the isospin diffusion data from NSCL/MSU have constrained the value of $x$
to be between $0$ and $-1$ for nuclear matter densities less than about $%
1.2\rho _{0}$ \cite{chen05,lichen05}, we will thus in the present work
consider the two values of $x=0$ and $x=-1$.

At zero temperature, $f_{\tau }(\vec{r},\vec{p})$ $=\frac{2}{h^{3}}\Theta
(p_{f}(\tau )-p)$ and the integral in Eqs.~(\ref{MDIV})~and (\ref{MDIU}) can
be calculated analytically \cite{das03}. For an asymmetric nuclear matter at
thermal equilibrium with a finite temperature $T$, the phase space
distribution function becomes the Fermi distribution
\begin{equation}
f_{\tau }(\vec{r},\vec{p})=\frac{2}{h^{3}}\frac{1}{\exp (\frac{\epsilon
(\rho ,T,\delta ,\vec{p},\tau )-\mu _{\tau }}{T})+1}  \label{f}
\end{equation}%
where $\mu _{\tau }$ is the chemical potential and $\epsilon (\rho ,T,\delta
,\vec{p},\tau )$ is the total single particle energy for a nucleon with
isospin $\tau $ and momentum $\vec{p}$, which includes the kinetic energy
and the single particle potential $U(\rho ,T,\delta ,\vec{p},\tau )$, i.e.,
\begin{eqnarray}
\epsilon (\rho ,T,\delta ,\vec{p},\tau ) &=&\frac{p^{2}}{2m_{\tau }}+A_{u}%
\frac{\rho _{-\tau }}{\rho _{0}}+A_{l}\frac{\rho _{\tau }}{\rho _{0}}  \notag
\\
&+&B(\frac{\rho }{\rho _{0}})^{\sigma }(1-x\delta ^{2})-8\tau x\frac{B}{%
\sigma +1}\frac{\rho ^{\sigma -1}}{\rho _{0}^{\sigma }}\delta \rho _{-\tau }
\notag \\
&+&R_{\tau ,\tau }(\rho ,\vec{p})+R_{\tau ,-\tau }(\rho ,\vec{p})
\label{mdie}
\end{eqnarray}%
with%
\begin{equation}
R_{\tau _{1},\tau _{2}}(\rho ,\vec{p})=\frac{2C_{\tau _{1},\tau _{2}}}{\rho
_{0}}\int d^{3}p^{\prime }\frac{f_{\tau _{2}}(\vec{r},\vec{p}^{\prime })}{1+(%
\vec{p}-\vec{p}^{\prime })^{2}/\Lambda ^{2}}  \label{R}
\end{equation}%
where $\tau _{1}$ and $\tau _{2}$ can be chosen as the same or different to
mimic the last two terms of Eq. (\ref{MDIU}).

The chemical potential $\mu _{\tau }$ is therefore independent of the
nucleon momentum $\vec{p}$ and can be determined from
\begin{equation}
\rho _{\tau }=\frac{8\pi }{h^{3}}{\int_{0}^{\infty }}\frac{p^{2}dp}{\exp (%
\frac{\epsilon (\rho ,T,\delta ,\vec{p},\tau )-\mu _{\tau }}{T})+1}{.}
\label{mu}
\end{equation}%
From Eq.~(\ref{mu}), we can see that, at finite temperature, to obtain the
chemical potential $\mu _{\tau }$ requires knowing $\epsilon _{(}\rho
,T,\delta ,\vec{p},\tau )$ (and thus $R_{\tau _{1},\tau _{2}}(\rho ,\vec{p})$%
) for all $\vec{p}$, while from Eq.~(\ref{R}) knowing $R_{\tau _{1},\tau
_{2}}(\rho ,\vec{p})$ needs further to know $f_{\tau }(\vec{r},\vec{p})$
which again depends on the chemical potential $\mu _{\tau }$ from Eq.~(\ref%
{f}). Therefore. Eqs. (\ref{f}), (\ref{mdie}), (\ref{R}), and (\ref{mu})
constitute closed sets of equations whose solution can be obtained by a
self-consistency iteration, just as in the Hartree-Fock theory.

Following the recipe used in Ref.~\cite{gale}, the self-consistency problem
of Eqs. (\ref{f}), (\ref{mdie}), (\ref{R}), and (\ref{mu}) can be solved by
the following iterative scheme. Firstly, we make an initial guess for $%
R_{\tau _{1},\tau _{2}}(\rho ,\vec{p})$ from the zero temperature condition,
i.e.,
\begin{equation}
R_{\tau _{1},\tau _{2}}^{0}(\rho ,\vec{p})=\frac{2C_{\tau _{1},\tau _{2}}}{%
\rho _{0}}\int d^{3}p^{\prime }\frac{\frac{2}{h^{3}}\Theta (p_{f}(\tau
_{2})-p^{\prime })}{1+(\vec{p}-\vec{p}^{\prime })^{2}/\Lambda ^{2}}
\label{R0}
\end{equation}%
where $p_{f}(\tau )=\hbar (3\pi ^{2}\rho _{\tau })^{1/3}$ is the Fermi
momentum. The right hand side of Eq.~(\ref{R0}) is not related to the
chemical potential $\mu _{\tau }$ and thus the initial form of the single
nucleon energy $\epsilon ^{0}(\rho ,T,\delta ,\vec{p},\tau )$ can be
obtained. Secondly, substitute $\epsilon ^{0}(\rho ,T,\delta ,\vec{p},\tau )$
into Eq.~(\ref{mu}) to obtain the initial chemical potential $\mu _{\tau
}^{0}$ for protons and neutrons. Then, use $\epsilon ^{0}(\rho ,T,\delta ,%
\vec{p},\tau )$ and $\mu _{\tau }^{0}$ to obtain new $R_{\tau _{1},\tau
_{2}}(\rho ,\vec{p})$ function, namely, $R_{\tau _{1},\tau _{2}}^{1}(\rho ,%
\vec{p})$ from Eq.~(\ref{f}) and (\ref{R}). This in turn gives the new
single nucleon energy $\epsilon ^{1}(\rho ,T,\delta ,\vec{p},\tau )$ from
Eq. (\ref{mdie}) and then the new chemical potential $\mu _{\tau }^{1}$ can
be obtained from Eq.~(\ref{mu}). The cycle is repeated and a few iterations
are sufficient to achieve convergence for the chemical potential $\mu _{\tau
}$ with enough accuracy. It should be mentioned that the neutron and proton
chemical potentials are coupled with each other in asymmetric nuclear matter
and thus the convergence condition must be satisfied simultaneously for
neutrons and protons.

With the self-consistency iteration, we can finally obtain the chemical
potential $\mu _{\tau }$ and the single nucleon energy $\epsilon (\rho
,T,\delta ,\vec{p},\tau )$ for an asymmetric nuclear matter at thermal
equilibrium with a finite temperature $T$. The potential energy density $%
V(\rho ,T,\delta )$ of the thermal equilibrium asymmetric nuclear matter
then can be calculated from Eq.~(\ref{MDIV}) and the energy per nucleon $%
E(\rho ,T,\delta )$ is then obtained as

\begin{equation}
E(\rho ,T,\delta )=\frac{1}{\rho }\left[ V(\rho ,T,\delta )+{\sum_{\tau }}%
\int d^{3}p\frac{p^{2}}{2m_{\tau }}f_{\tau }(\vec{r},\vec{p})\right] .
\label{E}
\end{equation}%
Furthermore, we can obtain the entropy per nucleon $S_{\tau }(\rho ,T,\delta
)$ of the thermal equilibrium asymmetric nuclear matter as
\begin{equation}
S_{\tau }(\rho ,T,\delta )=-\frac{8\pi }{{\rho }h^{3}}\int_{0}^{\infty
}p^{2}[n_{\tau }\ln n_{\tau }+(1-n_{\tau })\ln (1-n_{\tau })]dp  \label{S}
\end{equation}%
with the occupation probability%
\begin{equation}
n_{\tau }=\frac{1}{\exp (\frac{\epsilon (\rho ,T,\delta ,\vec{p},\tau )-\mu
_{\tau }}{T})+1}.
\end{equation}%
Finally, the free energy per nucleon $F(\rho ,T,\delta )$ of the thermal
equilibrium asymmetric nuclear matter can be obtained from the thermodynamic
relation

\begin{equation}
F(\rho ,T,\delta )=E(\rho ,T,\delta )-T{\sum_{\tau }}S_{\tau }(\rho
,T,\delta ).  \label{F}
\end{equation}

\begin{figure}[tbh]
\includegraphics[scale=0.8]{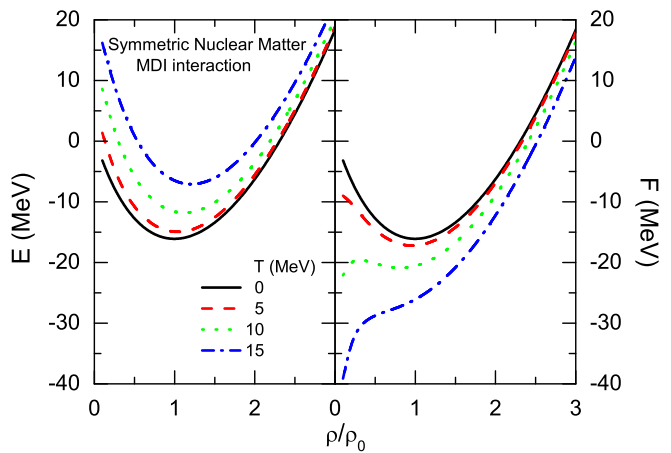}
\caption{{\protect\small (Color online) Density dependence of the energy per
nucleon }$E${\protect\small \ (left panel) and free energy per nucleon }$F$%
{\protect\small \ (right panel) for symmetric nuclear matter at }$T=0$%
{\protect\small \ MeV, }$5${\protect\small \ MeV, }$10${\protect\small \ MeV
and }$15${\protect\small \ MeV with the MDI interaction.}}
\label{fig1}
\end{figure}

Using the MDI interaction, we can now calculate the energy per nucleon $%
E(\rho ,T,\delta )$ and free energy per nucleon $F(\rho ,T,\delta )$ of
nuclear matter at finite temperature from Eq.~(\ref{E}) and (\ref{F}). Shown
in Fig. \ref{fig1} is the density dependence of $E(\rho ,T,\delta )$ and $%
F(\rho ,T,\delta )$ for symmetric nuclear matter at $T=0$ MeV, $5$ MeV, $10$
MeV and $15$ MeV using the MDI interaction with $x=0$ and $-1$. For
symmetric nuclear matter ($\delta =0$), the parameter $x=0$ would give the
same results as the parameter $x=$ $-1$ as we have discussed above, and thus
the curves shown in Fig. \ref{fig1} are the same for $x=0$ and $-1$. From
Fig. \ref{fig1}, one can see that the energy per nucleon $E(\rho ,T,\delta )$
increases with increasing temperature $T$ while the free energy per nucleon $%
F(\rho ,T,\delta )$ decreases with increment of $T$. The increment of the
energy per nucleon $E(\rho ,T,\delta )$ with the temperature reflects the
thermal excitation of the nuclear matter due to the change of the
phase-space distribution function $f_{\tau }(\vec{r},\vec{p})$. With the
increment of the temperature, more nucleons move to higher momentum states
and thus lead to larger internal energy per nucleon. On the other hand, the
decrement of the free energy per nucleon $F(\rho ,T,\delta )$ with $T$ is
mainly due to the increment of the entropy per nucleon with increasing
temperature. This feature also implies that the increment of $TS(\rho ,T)$
with $T$ is larger than the increment of $E(\rho ,T)$ with $T$. Furthermore,
the temperature effects are seen to be stronger at lower densities while
they become much weaker at higher densities. At lower densities, the Fermi
momentum $p_{f}(\tau )$ is smaller and thus temperature effects on the
energy per nucleon $E(\rho ,T,\delta )$ are expected to be stronger.
Meanwhile, the entropy per nucleon becomes larger at lower densities where
the particles become more free in phase space and thus leads to a smaller
free energy per nucleon.
\begin{figure}[tbh]
\includegraphics[scale=0.8]{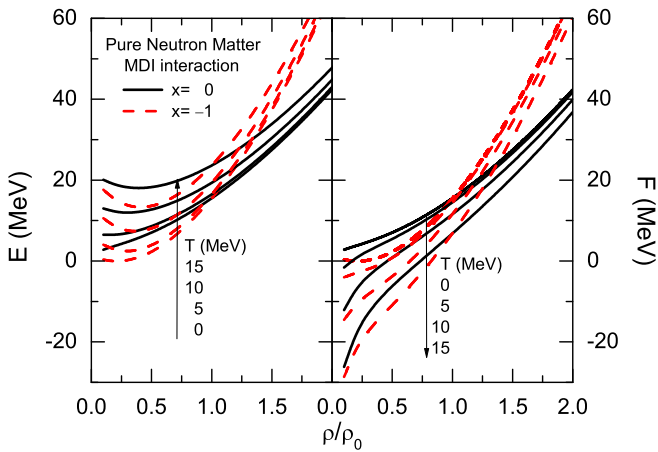}
\caption{{\protect\small (Color online) Same as Fig. \protect\ref{fig1} but
for pure neutron matter using the MDI interaction with }$x=0${\protect\small %
\ (solid lines) and }$-1${\protect\small \ (dashed lines).}}
\label{fig2}
\end{figure}

Similarly, shown in Fig. \ref{fig2} are the density dependence of the $%
E(\rho ,T,\delta )$ and $F(\rho ,T,\delta )$ for pure neutron matter at $T=0$
MeV, $5$ MeV, $10$ MeV and $15$ MeV using the MDI interaction with $x=0$ and
$-1$. The temperature dependence of the $E(\rho ,T,\delta )$ and $F(\rho
,T,\delta )$ for pure neutron matter is seen to be similar to that of the
symmetric nuclear matter as shown in Fig. \ref{fig1}. However, the
parameters $x=0$ and $-1$ display different density dependence for the
energy per nucleon $E(\rho ,T,\delta )$ and free energy per nucleon $F(\rho
,T,\delta )$, which just reflects that the parameters $x=0$ and $-1$ give
different density dependence of the nuclear symmetry energy and symmetry
free energy as will be discussed in the following.

\section{Nuclear symmetry energy and symmetry free energy}

\label{symmetry energy}
\begin{figure}[tbh]
\includegraphics[scale=0.8]{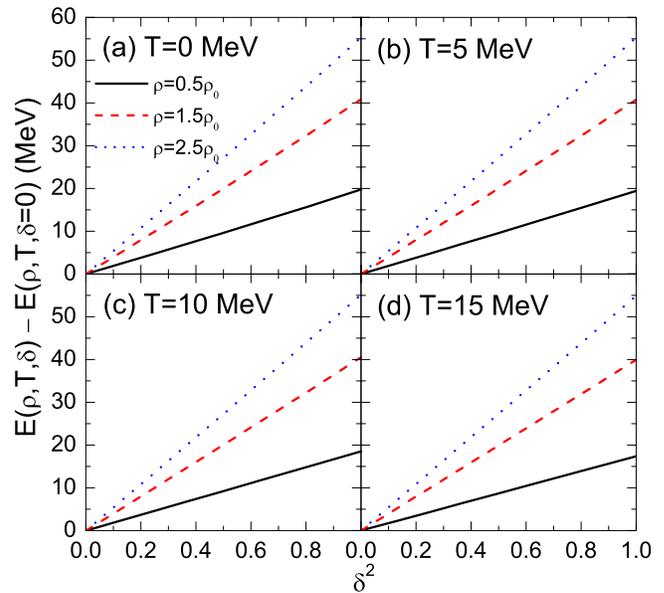}
\caption{{\protect\small (Color online) }$E(\protect\rho ,T,\protect\delta %
)-E(\protect\rho ,T,\protect\delta =0)${\protect\small \ as a function of }$%
\protect\delta ^{2}${\protect\small \ at temperature }$T=0${\protect\small \
MeV (a), }$5${\protect\small \ MeV (b), }$10${\protect\small \ MeV (c) and }$%
15${\protect\small \ MeV (d) for three different baryon number densities }$%
\protect\rho =0.5\protect\rho _{0},1.5\protect\rho _{0}${\protect\small \
and }$2.5\protect\rho _{0}${\protect\small \ using the MDI interaction with }%
$x=0${\protect\small .}}
\label{fig3}
\end{figure}
As in the case of zero temperature, phenomenological and microscopic studies
\cite{chen01,zuo03} indicate that the equation of state of hot neutron-rich
matter at density $\rho $, temperature $T$, and an isospin asymmetry $\delta
$ can also be written as a parabolic function of $\delta $, i.e.,
\begin{equation}
E(\rho ,T,\delta )=E(\rho ,T,\delta =0)+E_{sym}(\rho ,T)\delta ^{2}+\mathcal{%
O}(\delta ^{4}).  \label{eos}
\end{equation}%
The temperature and density dependent symmetry energy $E_{sym}(\rho ,T)$ for
hot neutron-rich matter can thus be extracted from $E_{sym}(\rho ,T)\approx
E(\rho ,T,\delta =1)-E(\rho ,T,\delta =0)$. The symmetry energy $%
E_{sym}(\rho ,T)$ is the energy cost to convert all protons in symmetry
matter to neutrons at the fixed temperature $T$ and density $\rho $. In
order to check the empirical parabolic law Eq. (\ref{eos}) for the MDI
interaction, we show in Fig. \ref{fig3} $E(\rho ,T,\delta )-E(\rho ,T,\delta
=0)$ as a function of $\delta ^{2}$ at temperature $T=0$ MeV, $5$ MeV, $10$
MeV and $15$ MeV for three different baryon number densities $\rho =0.5\rho
_{0},1.5\rho _{0}$ and $2.5\rho _{0}$ using the MDI interaction with $x=0$.
The clear linear relation between $E(\rho ,T,\delta )-E(\rho ,T,\delta =0)$
and $\delta ^{2}$ shown in Fig. \ref{fig3} indicates the validity of the
empirical parabolic law Eq. (\ref{eos}) for the hot neutron-rich matter. We
note that the empirical parabolic law Eq. (\ref{eos}) is also well satisfied
for the parameter $x=-1$.
\begin{figure}[tbh]
\includegraphics[scale=0.8]{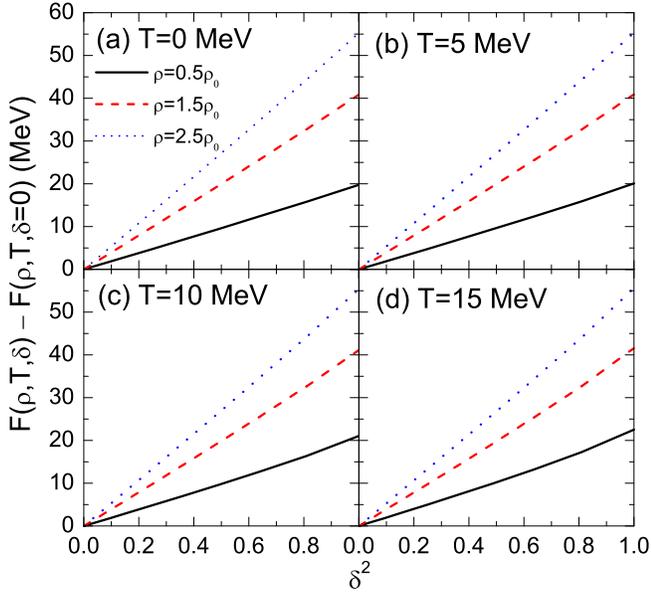}
\caption{{\protect\small (Color online) Same as Fig. \protect\ref{fig3} but
for the free energy per nucleon }$F(\protect\rho ,T,\protect\delta )$%
{\protect\small .}}
\label{fig4}
\end{figure}

Similarly, we can define the symmetry free energy $F_{sym}(\rho ,T)$ by the
following parabolic approximation to the free energy per nucleon
\begin{equation}
F(\rho ,T,\delta )=F(\rho ,T,\delta =0)+F_{sym}(\rho ,T)\delta ^{2}+\mathcal{%
O}(\delta ^{4}).  \label{eosF}
\end{equation}%
The temperature and density dependent symmetry free energy $F_{sym}(\rho ,T)$
for hot neutron-rich matter can thus be extracted from $F_{sym}(\rho
,T)\approx F(\rho ,T,\delta =1)-F(\rho ,T,\delta =0)$ which is just the free
energy cost to convert all protons in symmetry matter to neutrons at the
fixed temperature $T$ and density $\rho $. In order to check if the
empirical parabolic law is also valid for the free energy per nucleon of hot
neutron-rich matter, we show in Fig. \ref{fig4} $F(\rho ,T,\delta )-F(\rho
,T,\delta =0)$ as a function of $\delta ^{2}$ at temperature $T=0$ MeV, $5$
MeV, $10$ MeV and $15$ MeV for three different baryon number densities $\rho
=0.5\rho _{0},1.5\rho _{0}$ and $2.5\rho _{0}$ using the MDI interaction
with $x=0$. One can see from Fig. \ref{fig4} that the parabolic law Eq. (\ref%
{eosF}) is also approximately satisfied though at low densities and high
temperatures, the linear relation between $F(\rho ,T,\delta )-F(\rho
,T,\delta =0)$ and $\delta ^{2}$ is violated slightly. For the parameter $%
x=-1$, we also obtained the similar conclusion.
\begin{figure}[tbh]
\includegraphics[scale=0.8]{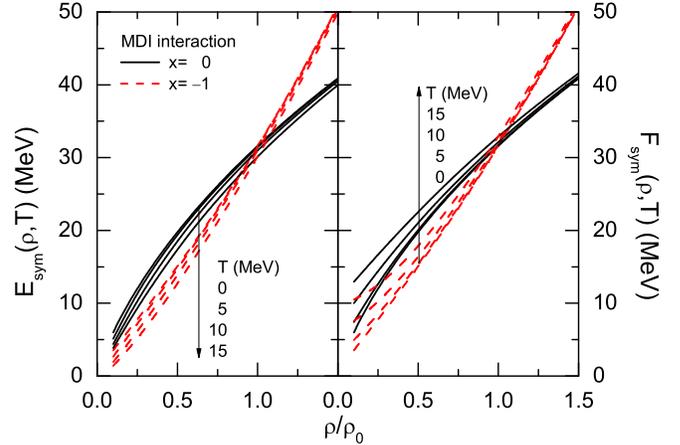}
\caption{{\protect\small (Color online) Density dependence of the symmetry
energy }$E_{sym}(\protect\rho ,T)${\protect\small \ (left panel) and the
symmetry free energy }$F_{sym}(\protect\rho ,T)${\protect\small \ (right
panel) at }$T=0${\protect\small \ MeV, }$5${\protect\small \ MeV, }$10$%
{\protect\small \ MeV and }$15${\protect\small \ MeV using the MDI
interaction with }$x=0${\protect\small \ and }$-1${\protect\small .}}
\label{fig5}
\end{figure}

In Fig. \ref{fig5}, we show the density dependence of the symmetry energy $%
E_{sym}(\rho ,T)$ and the symmetry free energy $F_{sym}(\rho ,T)$ at $T=0$
MeV, $5$ MeV, $10$ MeV and $15$ MeV using the MDI interaction with $x=0$ and
$-1$. For different choice of the parameter $x=0$ and $-1$, $E_{sym}(\rho
,T) $ and $F_{sym}(\rho ,T)$ display different density dependence with $x=0$
($-1 $) giving larger (smaller) values for the symmetry energy and the
symmetry free energy at lower densities while smaller (larger) ones at
higher densities for a fixed temperature. Similar to the $E(\rho ,T,\delta )$
and $F(\rho ,T,\delta )$ as shown in Figs. \ref{fig1} and \ref{fig2}, the
temperature effects on the symmetry energy $E_{sym}(\rho ,T)$ and the
symmetry free energy $F_{sym}(\rho ,T)$ are found to be stronger at lower
densities while they become much weaker at higher densities.

Interestingly, we can see from Fig. \ref{fig5} that the symmetry energy $%
E_{sym}(\rho ,T)$ and the symmetry free energy $F_{sym}(\rho ,T)$ exhibit
opposite temperature dependence, namely, with increasing temperature $T$, $%
E_{sym}(\rho ,T)$ decreases while $F_{sym}(\rho ,T)$ increases. This also
means that $F_{sym}(\rho ,T)$ always has larger values than $E_{sym}(\rho ,T)
$ at fixed density and temperature since they are identical at zero
temperature. At higher temperatures, one expects the symmetry energy $%
E_{sym}(\rho ,T)$ to decrease as the Pauli blocking (a pure quantum effect)
becomes less important when the nucleon Fermi surfaces become more diffused
at increasingly higher temperatures \cite{chen01,zuo03,lichen06EsymT}. On
the other hand, the symmetry free energy $F_{sym}(\rho ,T)$ is related to
the entropy per nucleon of the asymmetric nuclear matter, which is not a
pure quantum effect, and its increment with increasing temperature can be
understood by the following expression
\begin{eqnarray}
F_{sym}(\rho ,T) &=&E_{sym}(\rho ,T)  \notag \\
&&+T\left[ S_{n}(\rho ,T,\delta =0)+S_{p}(\rho ,T,\delta =0)\right]   \notag
\\
&&-TS_{n}(\rho ,T,\delta =1).  \label{Fsym}
\end{eqnarray}%
The first term of the right hand side in Eq. (\ref{Fsym}) is the
symmetry energy $E_{sym}(\rho ,T)$, which decreases with
increasing temperature as discussed above. However, the total
entropy per nucleon of the symmetric nuclear matter is larger than
that of the pure neutron matter and their difference becomes
larger with increasing temperature, which leads to a positive
value for the difference between the last two terms of the right
hand side in Eq. (\ref{Fsym}). Therefore, $F_{sym}(\rho ,T)$ has
larger values than $E_{sym}(\rho ,T)$ at fixed density and
temperature. Furthermore, the increment of $TS(\rho ,T)$ with $T$
is stronger than the increment of $E(\rho ,T)$ with $T$ as
mentioned above, and the combinational effects thus cause the
symmetry free energy $F_{sym}(\rho ,T)$ increase with increasing
temperature.

\begin{figure}[tbh]
\includegraphics[scale=1.2]{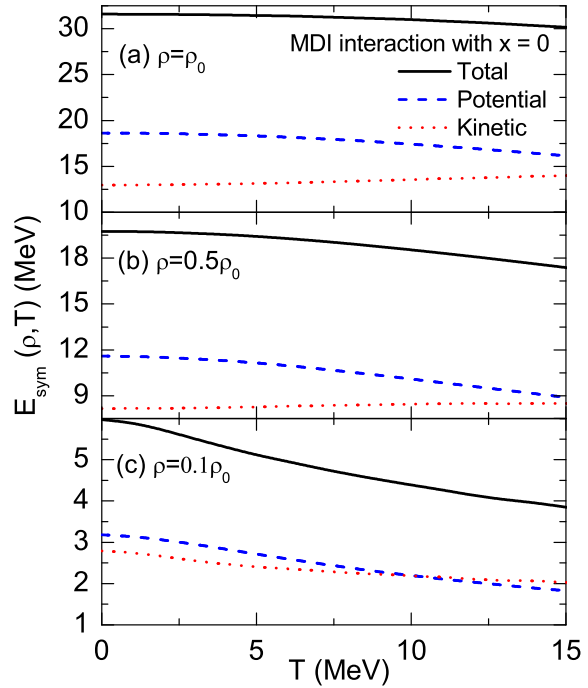}
\caption{{\protect\small (Color online) Temperature dependence of
the symmetry energy $E_{sym}(\protect\rho ,T)$ as well as its
potential energy
part and kinetic energy part using MDI interaction with $x=0$ at $\protect%
\rho =\protect\rho _{0}$ (a), $0.5\protect\rho _{0}$ (b), and $0.1\protect%
\rho _{0}$ (c).}} \label{EsymTPotKinX0}
\end{figure}

Within the present self-consistent thermal model, because the single
particle potential is momentum dependent with the MDI interaction, the
potential part of the symmetry energy is expected to be temperature
dependent. It is thus interesting to study how the potential and kinetic
parts of the symmetry energy $E_{sym}(\rho ,T)$ may vary respectively with
temperature. However, for the symmetry free energy $F_{sym}(\rho ,T)$, one
cannot separate its potential and kinetic parts since $F_{sym}(\rho ,T)$
depends on the entropy that is determined by the phase space distribution
function. Fig.~\ref{EsymTPotKinX0} displays the temperature dependence of
the symmetry energy $E_{sym}(\rho ,T)$ as well as its potential and kinetic
energy parts using the MDI interaction with $x=0$ at $\rho =\rho _{0}$, $%
0.5\rho _{0}$, and $0.1\rho _{0}$. With the parameter $x=-1$, the same
conclusion is obtained. It is seen that both the symmetry energy $%
E_{sym}(\rho ,T)$ and its potential energy part decrease with
increasing temperature at all three densities considered. While
the kinetic energy part of the $E_{sym}(\rho ,T)$ increases
slightly with increasing temperature for $\rho =\rho _{0}$ and
$0.5\rho _{0}$ and decreases for $\rho =0.1\rho _{0}$. These
features are uniquely determined by the momentum dependence in the
MDI interaction within the present self-consistent thermal model.
The decrement of the kinetic energy part of the symmetry energy
with temperature at very low densities is consistent with
predictions of the Fermi gas model at high temperatures and/or
very low densities \cite{lichen06EsymT,mlz}. In the study of Ref.\
\cite{lichen06EsymT} using the simplified degenerate Fermi gas
model the potential part of the symmetry energy was assumed to be
temperature independent for simplicity. The decrease of the
symmetry energy observed there is thus completely due to the
decrease in the kinetic contribution. However, we note that the
temperature dependence of the total symmetry energy $E_{sym}(\rho
,T)$ is consistent with each other for the two models. This is due
to the fact that the phase space distribution function will vary
self-consistently according to if the single particle potential is
or not momentum dependent. From the present self-consistent
thermal model with momentum dependent MDI interaction, our results
indicate that the decreasing symmetry energy with increasing
temperature is essentially due to the decrement of its potential
contribution.

\section{Isotopic scaling in heavy-ion collisions}

\label{isoscaling}

It has been observed experimentally and theoretically in many types of
reactions that the ratio $R_{21}(N,Z)$ of yields of a fragment with proton
number $Z$ and neutron number $N$ from two reactions reaching about the same
temperature $T$ satisfies an exponential relationship $R_{21}(N,Z)\propto
\exp (\alpha N)$ \cite%
{betty01,shetty,sjy,shetty06,indra,wolfgang,kowalski,tsang01,botvina,ono,dorso,ma}%
. Particularly, in several statistical and dynamical models under some
assumptions \cite{tsang01,botvina,ono}, it has been shown that the scaling
coefficient $\alpha $ is related to the symmetry energy $C_{sym}(\rho ,T)$
via
\begin{equation}
\alpha =\frac{4C_{sym}(\rho ,T)}{T}\bigtriangleup \lbrack (Z/A)^{2}],
\label{scaling}
\end{equation}%
where $\bigtriangleup \lbrack (Z/A)^{2}]\equiv
(Z_{1}/A_{1})^{2}-(Z_{2}/A_{2})^{2}$ is the difference between the $%
(Z/A)^{2} $ values of the two fragmenting sources created in the two
reactions.

As mentioned in Ref. \cite{lichen06EsymT}, however, because of the different
assumptions used in the various derivations, the validity of Eq. (\ref%
{scaling}) is still disputable as to whether and when the
$C_{sym}$ is actually the symmetry energy or the symmetry free
energy. Moreover, the physical interpretation of the $C_{sym}(\rho
,T)$ is also not clear, sometimes even contradictory, in the
literature. The main issue is whether the $C_{sym}$ measures the
symmetry energy of the fragmenting source or that of the fragments
formed at freeze-out. This ambiguity is also due to the fact that
the derivation of Eq. (\ref{scaling}) is not unique. In
particular, within the grand canonical statistical model for
multifragmentation \cite{tsang01,botvina} the $C_{sym}$ refers to
the symmetry energy of primary fragments. While within the
sequential Weisskopf model in the grand canonical limit
\cite{tsang01} it refers to the symmetry
energy of the emission source. Following the arguments in Ref. \cite%
{lichen06EsymT}, we assume in the present work that the $C_{sym}$ reflects
the symmetry energy of \emph{bulk nuclear matter} for the emission source.
\begin{figure}[tbh]
\includegraphics[scale=0.85]{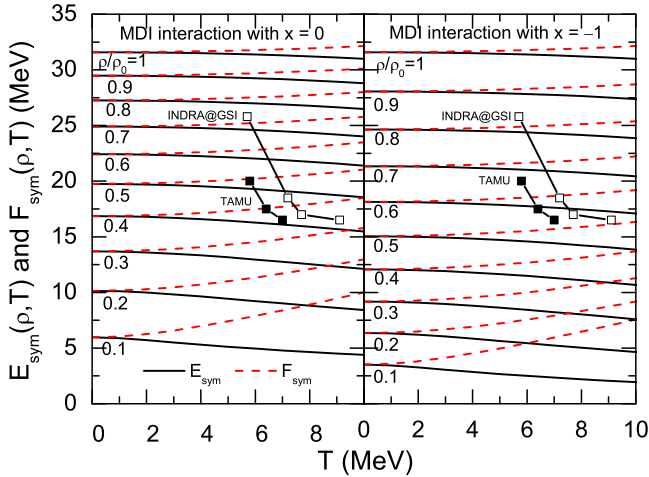}
\caption{{\protect\small (Color online) Temperature dependence of the
symmetry energy (solid lines) and symmetry free energy (dashed lines) using
MDI interaction with }$x=0${\protect\small \ (left panel) and }$-1$%
{\protect\small \ (right panel) at different densities from }$0.1\protect%
\rho _{0}${\protect\small \ to }$\protect\rho _{0}${\protect\small . The
experimental data from Texas A}$\&${\protect\small M University (solid
squares) and the INDRA-ALADIN collaboration at GSI (open squares) are
included for comparison.}}
\label{fig6}
\end{figure}

In Fig.~\ref{fig6}, we show the symmetry energy $E_{sym}(\rho ,T)$ and
symmetry free energy $F_{sym}(\rho ,T)$ as a function of temperature using
MDI interaction with $x=0$ and $-1$ at different densities from $0.1\rho
_{0} $ to $\rho _{0}$. The temperature dependence of the symmetry energy $%
E_{sym}(\rho ,T)$ is seen to be very similar to that in Ref.~\cite%
{lichen06EsymT} where a simplified degenerate Fermi gas model at finite
temperatures has been used. The symmetry energy does not change much with
the temperature at a given density, especially around the saturation density
$\rho _{0}$. Furthermore, it is seen that while the symmetry energy $%
E_{sym}(\rho ,T)$\ deceases slightly with the increasing temperature at a
given density, the symmetry free energy $F_{sym}(\rho ,T)$ increases\
instead. Around the saturation density $\rho _{0}$, it is found that the
difference between the symmetry energy $E_{sym}(\rho ,T)$ and the symmetry
free energy $F_{sym}(\rho ,T)$ is quite small, i.e., only several percents,
though at higher temperature, compared with their values at $T=0$ MeV. This
feature confirms the assumption on identifying $C_{sym}(\rho ,T)$ to $%
E_{sym}(\rho ,T)$ at lower temperatures and not so low densities \cite%
{shetty,sjy,shetty06}. At low densities, on the other hand, the
symmetry free energy $F_{sym}(\rho ,T)$ exhibits a stronger
temperature dependence and it is significantly larger than the
symmetry energy $E_{sym}(\rho ,T)$ at moderate and high
temperatures. This is due to the fact that the entropy
contribution to the symmetry free energy $F_{sym}(\rho ,T)$
becomes stronger at low densities as mentioned in Sec. \ref{EOS}
and Sec. \ref{symmetry energy}. It should be noted that, at low
densities the entropy may be affected strongly by the clustering
effects \cite{horowitz,kowalski} which are not included in the
present work.

Experimentally, the temperature $T$ and scaling coefficient $\alpha $ (thus
the $C_{sym}$) of the fragment emission source can be directly measured
while the determination of the density of emission source usually depends on
the model used. Also included in Fig.~\ref{fig6} are the experimental data
of the measured temperature dependent symmetry energy from Texas A\&M
University (TAMU) (solid squares) \cite{shetty06} and the INDRA-ALADIN
collaboration at GSI (open squares) \cite{indra,wolfgang}. From Fig. \ref%
{fig6}, it is seen clearly that the experimentally observed evolution of the
symmetry energy is mainly due to the change in density rather than
temperature, as shown in Ref.~\cite{lichen06EsymT}. Meanwhile, we can
estimate from Fig. \ref{fig6} the average freeze-out density of the fragment
emission source from the measured temperature dependent symmetry energy
based on the isotopic scaling analysis in heavy-ion collisions. In
particular, using the symmetry energy $E_{sym}(\rho ,T)$ from the MDI
interaction with $x=0$, we find the average freeze-out density of the
fragment emission source $\rho _{f}$ is between about $0.41\rho _{0}$ and $%
0.52\rho _{0}$ for TAMU data while about $0.42\rho _{0}$ and $0.75\rho _{0}$
for INDRA-ALADIN collaboration data. On the other hand, using the symmetry
energy $E_{sym}(\rho ,T)$ from the MDI interaction with $x=-1$, the $\rho
_{f}$ is found to be between about $0.57\rho _{0}$ and $0.68\rho _{0}$ for
TAMU data while about $0.58\rho _{0}$ and $0.84\rho _{0}$ for INDRA-ALADIN
collaboration data. It is interesting to see that the extracted values of $%
\rho _{f}$ from the MDI interaction with $x=0$ is very similar to those
extracted in Ref. \cite{shetty06} using different models.

Furthermore, if the symmetry free energy $F_{sym}(\rho ,T)$ from the MDI
interaction with $x=0$ is used to estimate the $\rho _{f}$, we find the $%
\rho _{f}$ is between about $0.36\rho _{0}$ and $0.49\rho _{0}$ for TAMU
data and about $0.33\rho _{0}$ and $0.72\rho _{0}$ for INDRA-ALADIN
collaboration data. While if the symmetry free energy $F_{sym}(\rho ,T)$
from the MDI interaction with $x=-1$ is used, the $\rho _{f}$ is between
about $0.52\rho _{0}$ and $0.66\rho _{0}$ for TAMU data and about $0.51\rho
_{0}$ and $0.83\rho _{0}$ for INDRA-ALADIN collaboration data. Therefore,
the extracted $\rho _{f}$ values are not sensitive to if the measured $%
C_{sym}(\rho ,T)$ is the symmetry energy or the symmetry free energy.
However, the extracted $\rho _{f}$ values are indeed sensitive to the $x$
parameter used in the MDI interaction, namely, the density dependence of the
symmetry energy. We note that the zero-temperature symmetry energy for the
MDI interaction with $x=0$ and $-1$ can be parameterized, respectively, as $%
31.6(\rho /\rho _{0})^{0.69}$ MeV and $31.6(\rho /\rho _{0})^{1.05}$ MeV
\cite{chen05}. Therefore, the isotopic scaling in heavy-ion collisions
provides a potential good probe for the density dependence of the nuclear
matter symmetry energy once the average density of the emission source has
been determined in the isotopic scaling measurement, as pointed out in Ref.~%
\cite{lichen06EsymT}.

\section{Summary}

\label{summary}

Within a self-consistent thermal model using the isospin and momentum
dependent MDI interaction with $x=0$ and $-1$ constrained by the isospin
diffusion data in heavy-ion collisions, we have investigated the temperature
dependence of the nuclear matter symmetry energy $E_{sym}(\rho ,T)$ and
symmetry free energy $F_{sym}(\rho ,T)$. It is shown that the nuclear matter
symmetry energy $E_{sym}(\rho ,T)$ generally decreases with increasing
temperature while the symmetry free energy $F_{sym}(\rho ,T)$ exhibits
opposite temperature dependence. The decrement of the symmetry energy with
temperature is essentially due to the decrement of the potential energy part
of the symmetry energy with temperature. The temperature effects on the
nuclear matter symmetry energy and symmetry free energy are found to be
stronger at lower densities while become much weaker at higher densities.
Furthermore, the difference between the nuclear matter symmetry energy $%
E_{sym}(\rho ,T)$ and symmetry free energy $F_{sym}(\rho ,T)$ is found to be
quite small around nuclear saturation density, although significantly large
at very low densities.

Comparing the theoretical density and temperature dependent symmetry energy $%
E_{sym}(\rho ,T)$ with the $C_{sym}(\rho ,T)$ parameter extracted from the
isotopic scaling data from TAMU and the INDRA-ALADIN collaboration at GSI,
we found that the experimentally observed evolution of the symmetry energy
is mainly due to the change in density rather than temperature, as shown in
the previous work \cite{lichen06EsymT}. Meanwhile, we have estimated the
average freeze-out density of the fragment emission source formed in these
reactions by comparing the calculated $E_{sym}(\rho ,T)$ or $F_{sym}(\rho
,T) $ with the measured $C_{sym}(\rho ,T)$. Our results indicate that the
extracted average freeze-out densities are not sensitive to whether the
experimentally measured $C_{sym}(\rho ,T)$ parameter is the symmetry energy $%
E_{sym}(\rho ,T)$ or the symmetry free energy $F_{sym}(\rho ,T)$ in the
temperature and density ranges reached in the TAMU and INDRA/GSI
experiments. They are, however, sensitive to the $x$ parameter used in the
MDI interaction, namely, the density dependence of the symmetry energy.
Therefore the isotopic scaling in heavy-ion collisions provides a
potentially good probe for the density dependence of the nuclear matter
symmetry energy provided the average density of the emission source can be
determined simultaneously in the isotopic scaling measurements.

\begin{acknowledgments}
This work was supported in part by the National Natural Science Foundation
of China under Grant Nos. 10334020, 10575071, and 10675082, MOE of China
under project NCET-05-0392, Shanghai Rising-Star Program under Grant No.
06QA14024, the SRF for ROCS, SEM of China, the US National Science
Foundation under Grant Nos. PHY-0652548 and PHY-0456890, and the
NASA-Arkansas Space Grants Consortium Award ASU15154.
\end{acknowledgments}

\end{document}